# Lattice dynamics of solid cubane within the quasi-harmonic approximation

## Taner Yildirim*

*NIST Center for Neutron Research, National Institute of Standards and Technology, Building 235, Room E112, Gaithersburg, MD 20899, USA*

### Abstract

Solid cubane, which is composed of weakly interacting cubic molecules, exhibits many unusual and interesting properties, such as a very large thermal expansion and a first-order phase transition at $T_p = 394$ K from an orientationally ordered phase of $R\bar{3}$ symmetry to a *non-cubic* disordered phase of the same symmetry with a volume expansion of 5.4%, among the largest ever observed. We study the lattice dynamics of solid cubane within the quasi-harmonic and rigid-molecule approximation to explain some of these unusual dynamical properties. The calculated phonon density of states, dispersion curves and thermal expansion agree surprisingly well with available experimental data. We find that the amplitude of thermally excited orientational excitations (i.e. librons) increases rapidly with increasing the temperature and reaches about 35° just before the orientational phase transition. Hence, the transition is driven by large amplitude collective motions of the cubane molecules. Similarly the amplitude of the translational excitations shows a strong temperature dependence and reaches one-tenth of the lattice constant at $T = 440$ K. This temperature is in fair agreement with the experimental melting temperature of 405 K, indicating that the Lindemann criterion works well even for this unusual molecular solid.
Published by Elsevier Science Ltd.

## 1. Introduction

Cubane ($C_8H_8$), an atomic-scale cube, is an immensely strained molecule whose 90° bond angle challenges classical notions of the sp³ bonding of carbon [1–3]. Its high heat of formation [4], its exceptional symmetry, and the ease with which substitutions can be made on its skeleton make cubane an attractive template for a new generation of fuels and explosives [5]. Equally remarkable is the fact that cubane forms a stable solid at room temperature, with a crystalline structure composed of cubane molecules occupying the corners of a rhombohedral primitive unit cell (space group $R\bar{3}$) (see Fig. 1) [3]. The cubic molecular geometry gives the solid many unusual electronic, struc-tural, and dynamical properties compared to those of other hydrocarbons [6–15].

As in many molecular solids, solid cubane undergoes a first-order phase transition ($T_c = 394$ K) from an orientationally ordered phase to a non-cubic orientationally disordered (plastic) phase, resulting in a significant volume expansion of 5.4% [6,11]. Unlike the cubic plastic phases of many other molecular solids, this phase was also found to be rhombohedral. But it has a different rhombohedral angle, $\alpha = 103.3°$ [11]. The plastic phase persists until $T = 405$ K, at which point cubane melts. Compared to similar size hydrocarbons, the melting point of cubane is very high. The temperature dependence of the properties of solid cubane are also very interesting. It shows a very large thermal expansion, and recent quasi-elastic neutron scattering measurements indicate very large amplitude orientational dynamics [11,13]. All of these results indicate that solid cubane forms an ideal model system with which to challenge our understanding of molecular solids, orientational phase transition, and lattice dynamics.

* Tel.: +1-301-975-6228; fax: +1-301-921-9847.
  *E-mail address:* taner@nist.gov (T. Yildirim).

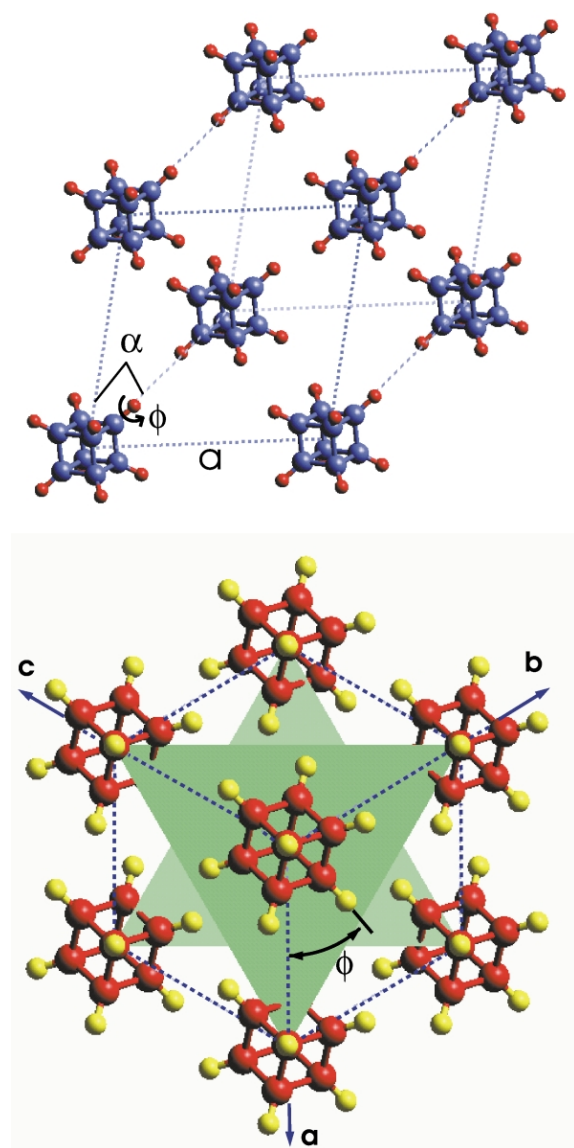

Fig. 1. Top: crystal structure of solid cubane. For clarity, the lattice constant is fixed at $a = 10$ Å. Large and small spheres represent the carbon and hydrogen atoms, respectively. Lattice parameters, $a$, $\alpha$, and $\phi$ are also depicted. Bottom: an oriented view of the crystal structure along the threefold [111] axis of the rhombohedral cell.

Recently a collaboration with Prof. Ciraci, to whom this issue is dedicated in honour of his 60th birthday, was began to study the novel electronic and structural properties of solid cubane and various cubane derivatives using DFT within the local density approximation (LDA) [15]. We found that many molecular properties such as the molecular orbitals, bond lengths and vibrational spectra can be accurately described by first-principles calculations. Sur-prisingly, even some of the solid state properties such as the lattice constants and orientation of the cubane molecules can be predicted reasonably well, even though the van der Waals interactions are not well described in LDA. However, all of these studies are limited to the static properties, and there has been little work done to explain the interesting dynamical properties of solid cubane theoretically. Here we address this issue by studying the lattice dynamics of solid cubane within the quasi-harmonic and rigid-molecule approximations.

This paper is organized as follows. In Section 2 we introduce a simple semi-empirical atom–atom type potential to describe the weak interaction between cubane molecules in the solid. In Section 3 we study the static energy of solid cubane as a function of three lattice parameters, namely the lattice constant $a$, the rhombohedral angle $\alpha$, the setting angle $\phi$ (see Fig. 1). The lattice dynamical analysis including phonon and libron dispersion curves and density of states (DOS) is given in Section 4. The temperature dependence of the structural parameters are discussed in Section 5. Here we also estimate the amplitude of thermally excited translational and orientational excitations. We then employ the Lindemann criterion [16] to estimate the melting temperature of cubane. Our conclusions are summarized in Section 6.

## 2. Intermolecular potential model

The intermolecular interactions between cubane molecules are described by a semi-empirical atom–atom potential of the form [17]:

$$V_{ij}(r_{ij}) = q_i q_j / r_{ij} + V_{ij}(\text{vdW}) + B_{ij} \exp(-C_{ij} r_{ij}) \tag{1}$$

where the first term represent the Coulomb interaction between molecules and the charges, $q_H = -q_C = 0.1$ are estimated from the calculated first-principles Mulliken charges. We observe that the Coulomb interaction does not contribute to the cohesive energy significantly but it is essential to obtain the right order of energy for the librational excitations. It tends to expand the lattice constant and to reduce the rhombohedral angle $\alpha$.

The second term in the above equation is the van der Waals (vdW) interactions between atoms $i$ and $j$ and it is estimated from the asymptotic limit by the London formula [17]

$$V_{i,j}(\text{vdW}) = -\frac{A_{ij}}{r_{ij}^6}, \tag{2}$$

where the parameters $A_{ij}$ are obtained using the Slater–Kirkwood approximation [18] as $A_{CC} = 16.066$ eV, $A_{CH} = 5.63$ eV, and $A_{HH} = 1.973$ eV.

Finally, the last term in the potential is to describe the short range repulsive interactions due to charge overlap between neighbor molecules. The exponents, $C_{HH}$, $C_{CH}$ and $C_{CC}$ were derived from several quantum mechanical calculations of the repulsion of two respective ions to be 3.74, 3.67, and 3.60 Å$^{-1}$, respectively, [17]. The other three parameters, $B_{HH}$, $B_{CH}$, and $B_{CC}$, were obtained from the equilibrium condition;

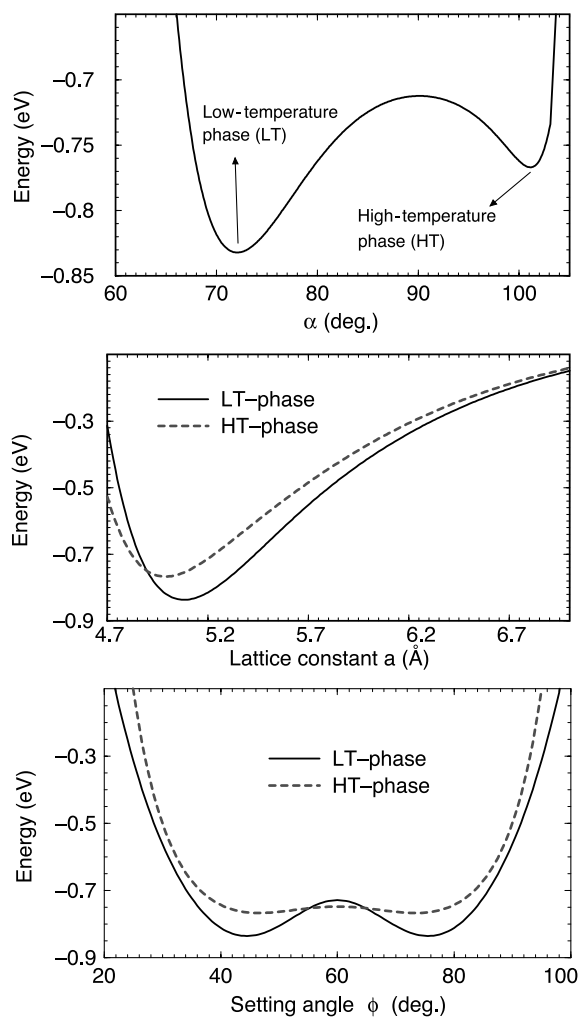

Fig. 2. Top: total energy of solid cubane as a function of the rhombohedral cell angle $\alpha$ with the lattice constant fixed at 5.1 Å, clearly illustrating the local and global minima corresponding to the experimentally observed low- and high-temperature phases. Middle: same as above except the lattice constant is varied with $\alpha$ fixed at values for LT- (solid line) and HT-phases (dashed line). Bottom: total energy as the molecules rotate about the three-fold axis of the crystal in the LT- (solid line) and HT-phases (dashed line).

$$B_{ij} = 6A_{ij}\frac{e^{C_{ij}r_{ij}^0}}{(r_{ij}^0)^7}C_{ij} \qquad (3)$$

where $r_{ij}^0$ are the standard set of van der Waals distances for atom $i$ and $j$. For C and H, we took $r_{HH}^0 = 2.6$ Å,

$r_{CH}^0 = 3.15$ Å, and $r_{CC}^0 = 3.7$ Å, respectively, [17]. From these values and from above equation, one obtains the coefficients: $B_{HH} = 65.867$ eV, $B_{CH} = 313.676$ eV, and $B_{CC} = 1718.49$ eV. We did not attempt to adjust these potential parameters so that the terms in the potential model given in Eq. (1) preserve their true physical meanings. As we will see below, this potential works pretty well for solid cubane.

## 3. Structure of solid cubane

Here we will briefly explain the rhombohedral structure of both low and high-temperature phases of solid cubane in terms of the static energy calculations using the above potential. The rhombohedral unit cell of solid cubane (see Fig. 1) can be viewed as an fcc lattice that has been squashed along one particular [111]-axis. Hence, the structure is characterized by three variables, namely the lattice constant $a$, the rhombohedral angle $\alpha$ (which would be 60° for an fcc structure), and the rotation angle of the molecule about the three-fold axis of the cell (see Fig. 1(b)). By varying these parameters, the energy of various possible structures, $E(a,\alpha,\phi)$, can be calculated. Fig. 2 shows the variation of the total energy as a function of one of these three variables while the other two are kept constant at their optimized values.

From the top panel in Fig. 2 it is apparent that the potential predicts one global and one local minimum. Remarkably these minima correspond to the observed low- and high-temperature phases. From this curve it is clear that the cubic structures (i.e. fcc for $\alpha = 60°$ and bcc for $\alpha = 109.47°$) have very high energies. The simple cubic (sc) structure ($\alpha = 90°$) is actually a local maximum. The global and local minima of the potential correspond to non-cubic structures with $\alpha \sim 72$ and $\sim 101.3°$, respectively, very close to the experimentally determined values, 72.7° and $\sim 103.3°$, respectively. The total energy at the global minimum is $-0.833$ eV, in excellent agreement with the experimental cohesive energy of $0.83 \pm 0.02$ eV [4].

The middle panel in Fig. 2 shows the lattice constant variation of the total energy as the other parameters are kept constant at their optimum values. The global minimum energy structure (solid line) has slightly larger lattice constant than the local minimum structure (dashed line). The energy minimum occurs at $a \approx 5.1$ Å, which is slightly smaller than the experimental lattice constant of 5.2 Å at 77 K.

Next we examined how the potential energy depends on the angular orientation of the molecule, which is related to the orientational excitations, or librons, of the solid. For this we calculate the potential energy as the molecules are rotated about their three-fold axis (the crystallographic [111] direction) for the global and local minimum energy structures, which correspond to the orientationally ordered and disordered phases, respectively. The bottom panel in Fig. 2 summarizes our results. In the ordered phase (solid line), the potential minimum occurs at $\phi = 44.5°$, which is in good agreement with the experimental value $\phi = 47.5°$.

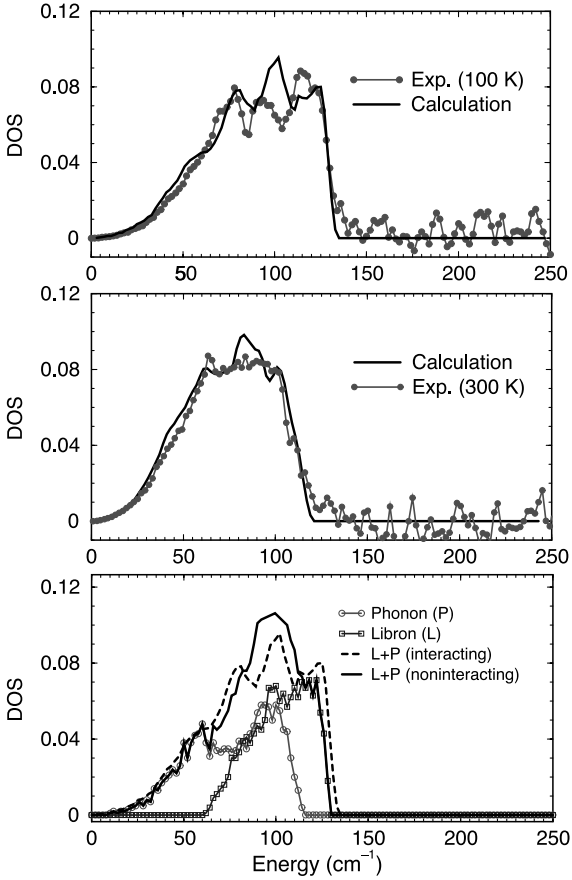

Fig. 3. Vibrational DOS obtained from inelastic neutron scattering [10] and calculated from a semi-empirical model potential at 100 K (top) and 300 K (middle). The bottom panel exhibits the contributions to DOS at 100 K from phonons and librons independently.

Perhaps the most interesting feature of our calculation is the finding that the potential energy as the molecules are rotated about the [111] axis in the disordered phase is very flat as shown in the bottom panel of Fig. 2 (dashed line). It is nearly independent of $\phi$ between 40 and 80°. The plastic phase of cubane therefore represents a nearly perfect example of a system with *collective large amplitude motions*, which suggests that the softening of the librational modes is the driving mechanism for the observed orientational phase transition. The collapse of the librations in the plastic phase will become more clear when we discuss the dispersion curves in Section 4.

## 4. Lattice dynamics of solid cubane

In this section we discuss the calculation of the vibrational (i.e. phonons) and orientational (i.e. librons) excitations in solid cubane with the quasi-harmonic approximation for perfectly rigid cubane molecules. The rigid-molecule approximation is well justified for cubane due to its very rigid cubic structure. To implement the rigid-molecule approximation, we introduce three translational coordinates as mass-weighted displacements of the center of mass along the three Cartesian axes by the relation

$$u_\rho^t(a) = M^{1/2} t_\rho(a) \qquad (4)$$

where $\rho = x, y, z$, $M$ is the mass of the cubane molecule, and $a$ is the sublattice index. Similarly, for the three degrees of rotational freedom, we define three rotational coordinates as inertia moment-weighted rotations around the molecular axes

$$u_\rho^r(a) = I^{1/2} \theta_\rho(a) \qquad (5)$$

where $I$ is the moment of inertia of cubane molecule.

After expanding the potential energy of a molecular crystal in terms of $u^t(a)$ and $u^r(a)$, the normal modes and frequencies are obtained by diagonalizing the dynamical interaction matrix

$$D(q)U(q) = \omega^2(q)U(q) \qquad (6)$$

where, the dynamical matrix is

$$D_{\alpha,\beta}^{i,i'}(q) = \frac{1}{\sqrt{m^i m^{i'}}} \sum_b \left[ \frac{\delta^2 V_I}{\delta u_\alpha^i(0)\delta u_\beta^{i'}(b)} \right]_0 e^{iq \cdot X(b)} \qquad (7)$$

where $V_I$ is the intermolecular potential and $m^i = M, I$ for $i = t, r$, respectively. The analytical expression for the derivative of $V_I$ for an atom–atom type potential as given in Eq. (1) can be found in Refs. [17,19]. In Ref. [19] this formalism has been successfully applied to the lattice dynamics studies of solid $C_{60}$. A complete descriptive review of the technique can also be found in Ref. [17]. The interacting libron–phonon spectrum is obtained by diagonalizing the full dynamical matrix $D_{\alpha,\beta}^{i,i'}(q)$, while the non-interacting libron and phonon spectrum is obtained from the $3 \times 3$ block-diagonal matrices (i.e. $i = i' = t$ for phonon and $i = i' = r$ for libron spectrum in $D_{\alpha,\beta}^{i,i'}(q)$).

Since in the $R\bar{3}$ structure of solid cubane there is only one molecule in the unit cell, there will be three translational modes (phonons) and three rotational modes (librons). The three phonons are the acoustic modes, corresponding to translation of the lattice as a whole at $\mathbf{q} = 0$ and thus cannot be observed by optical techniques. On the other hand, at $\mathbf{q} \approx 0$, librations are even under inversion and the group representation is $\Gamma_R = A_g + E_g$, indicating that they are Raman active. The $A_g$ mode corresponds to the libration of the molecules about the [111] axis, thus preserving the $R\bar{3}$ symmetry of the ground state. The $E_g$ modes are two-fold degenerate and correspond to librations about an axis perpendicular to [111]. Even though the librons are Raman active, their observation by Raman scattering is problematic in view of the isotropic polarizability tensor of the cubic molecule [8,9]. The inelastic neutron scattering technique, on the other hand, is not subject to any selection rules and therefore it is the best way to observe the lattice modes, not only at $\mathbf{q} \approx 0$ but also everywhere in the Brillouin zone [20]. However, due to lack of a single crystal of deuterated cubane sample, available experimental data are limited to the neutron measurements of the phonon DOS only [10,12, 13].

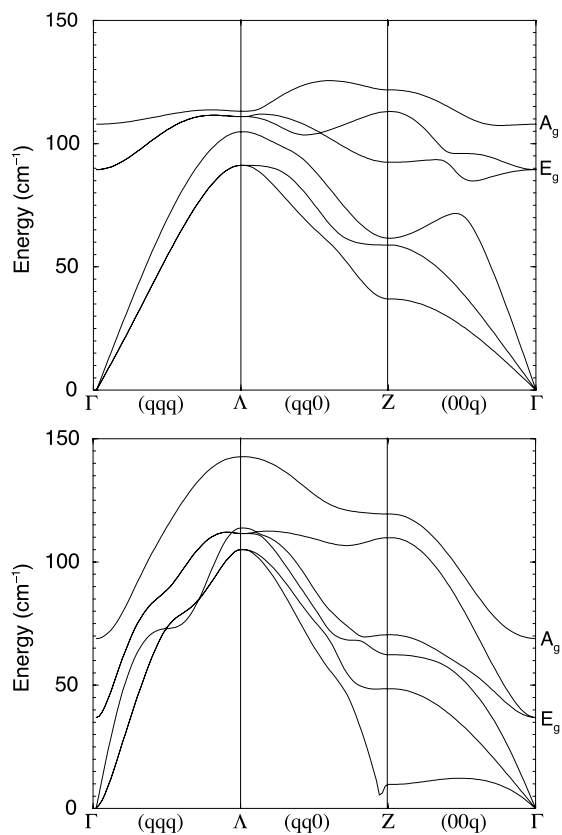

Fig. 4. The calculated phonon and libron dispersion curves along several high-symmetry directions in the Brillouin zone for both low-temperature (top) and high-temperature (bottom) phases of solid cubane. Note the significant softenning of the librons and one of the phonon branches near the Γ and Z-points, respectively, in the HT-phase.

Fig. 3 shows both the calculated phonon DOS and the measured generalized density of states (GDOS) [10] at two different temperatures $T = 100$ and 300 K. The agreement between calculation and experiment is quite good at both temperatures, considering the fact that we did not attempt to adjust the parameters in the potential and obtained the lattice parameters at 100 and 300 K self-consistently within quasi-harmonic approximation as discussed in Section 5. Both the experimental and calculated DOS indicate three main features centered at 78, 94, and 114 cm$^{-1}$. There is a significant softening of these features with increasing temperature, which is nicely reproduced by the calculations. Fits to a quadratic $q$ dependence indicate a low-frequency behavior that is Debye-like at both temperatures up to $\sim 60$ cm$^{-1}$. The bottom panel of Fig. 3 shows the contributions to the DOS from librons and phonons, indicating that librons lie well above the phonons in energy. The total DOS with and without libron–phonon interaction (solid and dashed lines, respectively) are very similar, indicating that there is a little coupling between librons and phonons. We also notice that below 50 cm$^{-1}$, no contribution to the DOS from the librational excitations. Similarly, above 110 cm$^{-1}$ the DOS is mainly due to the librational excitations. Both libron and phonon DOS exhibit a peak around 100 cm$^{-1}$. Hence we assign the highest two energy features observed at 94 and 114 cm$^{-1}$ to mostly the librational modes while the intensity at 78 cm$^{-1}$ corresponds to the two transverse-acoustic (TA) modes. Because the TA-mode dispersion flattens out at the zone boundary, the measured frequencies will be weighted most heavily by those in this region of the Brillouin zone. These interpretations of the neutron data are nicely confirmed by the calculated dispersion curves of the six external modes.

Fig. 4 shows the calculated libron and phonon dispersion curves along various symmetry directions in the Brillouin zone for both low- and high-temperature phases of solid cubane. In both phases the librons are clearly separated from the phonons with a little coupling. In the low-temperature phase (top), the librons are weakly momentum dependent, forming two main bands centered at 94 and 114 cm$^{-1}$ with mostly $E_g$ and $A_g$ symmetry, confirming the assignment of the modes discussed above. Unlike the situation in the low-temperature phase, the librons essentially collapse in the high-temperature phase as a result of the *particle-in-a-box-like* orientational potential, as discussed in the previous section. There are several clear indications of the instability of the system near the Γ- and Z-points. Since the potential is very anharmonic in the high-temperature phase, the lattice dynamical calculations probably break down for this phase. Therefore we will not discuss the high-temperature phase further in this paper.

## 5. Thermal expansion of solid cubane

Understanding the temperature dependence of the structural parameters of solid cubane is as important as understanding the orientational phase transition itself. Consequently we calculated the temperature dependence of the structural parameters within the quasi-harmonic approximation for solid cubane from zero temperature to well above the melting temperature. The results were

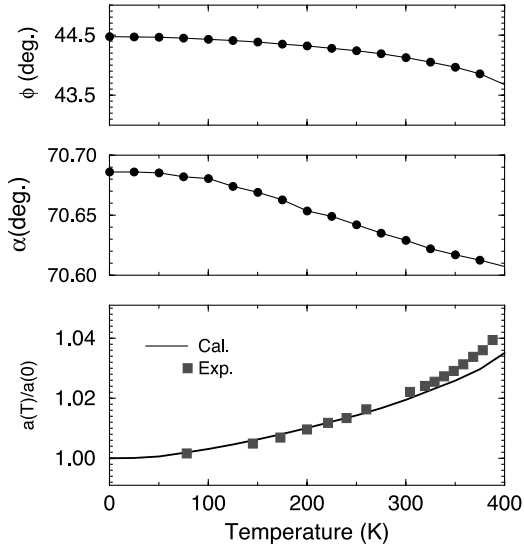

Fig. 5. The calculated temperature dependence of the structural parameters of solid cubane; the setting angle $\phi$ (top), the rhombohedral angle $\alpha$ (middle), and the thermal expansion of the lattice constant normalized to the zero temperature lattice constant (bottom). The squares shows the experimental data from Ref. [11].

obtained by minimizing the free energy, which is given as

$$F = V_l(a, \alpha, \phi) + \sum_j \sum_q \left[ \frac{1}{2} \hbar \omega_j(q) \right.$$
$$\left. + kT \ln(1 - \exp(-\hbar \omega_j(q)/kT)) \right] \tag{8}$$

where the first term is the static intermolecular potential energy and the second term is obtained by summing the six external modes over the wavevectors in the Brillouin zone. The $q$-summation is performed using $20 \times 20 \times 20$ k-points

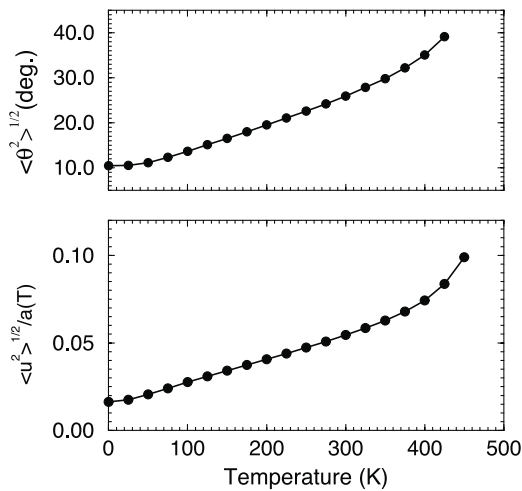

Fig. 6. The calculated root-mean-square (rms) displacements of the thermally populated orientational (top) and translational excitations.

according to the Monkhorst–Pack special k-point scheme [21].

In the strict harmonic approximation $F$ only depends on the lattice parameters through the static term $V_l$. In this case purely anharmonic crystal properties such as thermal expansion cannot be described at all. In the quasi-harmonic approximation the effect of the anharmonicity in the lattice energy is treated by allowing the frequencies of the phonons and librons to depend on the lattice parameters. Hence, for a given temperature $T$, we first take $a$ and $\alpha$, minimize the potential with respect to the molecular orientation (i.e. $\phi$), and then calculate the phonon and libron spectrum, and then the free energy. Repeating this for other values of $a$ and $\alpha$, we find the optimum values of the lattice parameters that minimize the free energy at a given temperature. In this way, one obtains the temperature dependence of the lattice parameters.

The calculated temperature dependence of the lattice parameters, $\phi$, $\alpha$, and $a$ is shown in Fig. 5. We note that $\alpha$ does not change within the experimental uncertainties until $T = T_p$, in perfect agreement with our calculations. Similarly, the setting angle of the cubane molecule, $\phi$, changes only weakly from 44.5° at 0 K to 43.5° at 400 K, also in agreement with the experimental observation. In contrast, the thermal expansion of the lattice constant $a$ varies by 4% between 0 K and $T_p = 400$ K. The bottom panel shows that the calculated total thermal expansion (solid line) is in excellent agreement with the experimental data (squares). The calculated curve lies slightly below the experimental data above $T \approx 350$ K, at which point we believe that the quasi-harmonic approximation starts to break down due to large amplitude vibrations. The large thermal expansion of the lattice constant can be understood as follows: when cubane is orientationally ordered, the molecules can move closer together, resulting in a smaller lattice constant. As the temperature increases, the amplitude of the rotational excitations increases, giving the molecules a larger effective radius (see Fig. 1(a)), causing the lattice to expand faster than in most other crystals.

Finally we discuss the root-mean-square (rms) values of the phonon and libron amplitudes as a function of temperature. The rms value of the phonon amplitude is important and can be used to estimate the melting point of a solid using the Lindemann criterion [16]. Also the temperature evolution of the libron amplitude may determine the validity of the Taylor expansion used in the lattice dynamics calculations.

The rms-values for the libron and phonon amplitudes are estimated from the formula;

$$\langle U^{i2} \rangle = \frac{1}{8N} \sum_q \sum_\tau \frac{\hbar}{m^i \omega_\tau(q)} \coth\left(\frac{\hbar \omega_\tau(q)}{2kT}\right) \tag{9}$$

where $i = t, r$ for phonon and libron, respectively. A detailed derivation of this formula can be found in Ref. [19]. The calculated rms-values of the phonon and libron amplitudes are shown in Fig. 6. We note that the libration amplitude is quite large due to the relatively small inertia of the cubane molecule. It increases with increasing temperature and reaches 30–40° above room temperature, indicating the presence of very large amplitude reorientational dynamics. Since the rms values are so large near 400 K, the quasi-harmonic theory probably break downs. This also explains the small deviation of the calculated thermal expansion from the experimental data at high-temperatures, as shown in the bottom panel of Fig. 5.

The temperature dependence of the rms-values of the translational motion of the cubane is shown in the bottom panel in Fig. 6. It increases roughly linearly with temperature and reaches 0.1 around 440 K. According to the Lindemann criterion [16], the solid should melt at this temperature. Indeed the melting point of cubane is 405 K, suggesting that the Lindemann criterion works fairly well even for such an unusual molecular solid.

## 6. Conclusion

We have presented a simple theory within the quasi-harmonic and rigid-molecule approximations to explain several unusual and interesting lattice dynamical properties of solid cubane. The calculated libron and phonon energies, DOS, and thermal expansion are in good agreement with available experimental data, indicating that our simple theory captures the essential physics in solid cubane. It is found that the correlated librational motions of the cubane molecules play a central role in controlling the temperature evolution of the structure. Despite many surprisingly good results obtained using the quasi-harmonic approximation, the theory seems to break down near the orientational phase transition temperature due to the large amplitude of the librational motions. Hence, it will be of interest to study the reorientational dynamics of cubane in the disordered phase using a technique where the harmonic approximation is not used. Studies based on molecular dynamics and mean-field theory will therefore be very useful. We are currently performing such studies, which will be published elsewhere.

## Acknowledgements

It is the author's pleasure to acknowledge many fruitful and interesting collaborations, including several on solid cubane and carbon nanotubes with Prof. Salim Ciraci, to whom this work is dedicated in honor of his 60th birthday.